\documentstyle[preprint,aps]{revtex}
\begin{document}
\draft
\title{\bf{\LARGE{ Stationary localized states due to quadratic nonlinearity
in one dimensional systems}}}
\author{Anandamohan Ghosh $^{\star}$, B C Gupta and K Kundu}
\address{Institute of Physics, Bhubaneswar - 751 005, India}
\address{ $^{\star}$ Permanent address: Department of Physics, University of 
Pune, Pune - 411007, India}
\maketitle
\begin{abstract}
 We investigate the effect of a nondegenerate quadratic nonlinear dimeric
impurity on the formation of stationary localized states in one dimensional
systems. We also consider the formation of stationary localized states in
a fully nonlinear system where alternative sites have same nonlinear 
strengths. Appropriate ansatzs have been chosen for all the cases which 
reproduce known results for special cases. Connection of the stability
of a state with its energy is presented graphically.

\end{abstract}

\narrowtext
\section{Introduction}

The discrete nonlinear Schr$\ddot{\rm o}$dinger equation which describes a
number of phenomena in condensed matter physics, non-linear optics and other
fields of physics [1-6] may be generally written as  
\begin{eqnarray}
i \frac{dC_{m}}{dt} &=& - \chi_{m} f_{m}(\mid C_{m} \mid) C_{m} 
+ V_{m,m+1} C_{m+1} 
+ V_{m,m-1} C_{m-1} \nonumber \\
{\rm where}~~~ V_{m,m+1} &=& V_{m+1,m}^{\star};~ {\rm and}
~m=1,2,3,...n.
\end{eqnarray}

In eq.(1) $\chi_{m}$ is the nonlinearity parameter associated with $m$-th
grid point and $V_{m,m+1}$ is the nearest neighbor hopping matrix element .
Since, $\sum_m |C_m|^2$ is made unity by choosing appropriate initial 
conditions, $|C_m|^2$ can be interpreted as the probability of finding a 
particle at the $m$-th grid point.
The nonlinearity term $f_{m}(\mid C_{m} \mid)$ arises due to the coupling of
the vibration of masses at the lattice points to the motion of a quasi-particle
in the high frequency lattice vibration limit [3]. The DNLSE in other
physical context is known as the discrete self trapping equation (DST). A
number of works on the DNLSE and the DST have been reported [7-10].

As for the application of the DNLSE particularly in condensed matter physics,
we cite among others the exciton propagation in Holstein molecular
crystal chain [2]. In general, the exciton propagation in quasi one 
dimensional systems [11]  having short range electron phonon 
interaction can
be adequately modeled by the DNLSE. Other examples include nonlinear 
optical responses in superlattices formed by dielectric or magnetic 
slabs [12]  and the mean field theory of a periodic array of 
twinning planes 
in the high $T_c$ superconductors [13]. We also note that 
the vibration in the nonlinear Klein-Gordon chain in certain restrictive
conditions can be described by the DNLSE [14]. 

One important feature of the DNLSE is that this can yield stationary 
localized ( SL ) states. These SL states might play a relevant role 
in the nonlinear DNA dynamics [15] and also in the energy localization in 
nonlinear lattices [16]. It has been shown that the presence of a 
nonlinear impurity can produce SL states in one, two and three 
dimension [17-24].  
The formation of SL states due 
to the presence of a single and a degenerate dimeric nonlinear impurity in 
a few linear hosts has been studied in details [20]. 
The same problem is also studied starting from an appropriate 
Hamiltonian [23]. The fixed point of the Hamiltonian 
[21-24] which generates the appropriate DNLSE can also produce the correct
equations governing the formation of SL states. We further note that 
the appropriate ansatz for the dimer problem has been obtained in our 
earlier analysis [20]. Furthermore, the formation of intersite peaked 
and dipped stationary localized states has been studied using the 
dimeric ansatz [24]. The formation of SL states in a perfect nonlinear chain
containing one nonlinear impurity as well as a degenerate nonlinear dimeric 
impurity has been studied [24].  In another study 
the formation of SL states in a perfect nonlinear Cayley tree and in a linear
Cayley tree with dimeric impurity has been considered.

 So far degenerate 
nonlinear dimer impurity has been considered . Naturally question arises what
happens if the impurities in the dimer are nondegenerate . We therefore plan
to study the effect of nondegeneracy in impurities on the formation of SL
states. Three linear hosts are studied, namely, a perfect one dimensional
chain , a Cayley tree and a linear chain with a bond defect .
We also consider the formation of SL states in a fully nonlinear chain having
a regular binary alloy composition. Appropriate ansatzs have been fitted 
in all the cases to obtain the results.

The organization of the paper is as follows. In sec. II we consider the
effect of nonlinear nondegenerate dimer on the formation of SL states 
in linear systems. Sec. III deals with the formation of SL states in a
fully nonlinear system and the stability analysis of the SL states.
Finally we summarize our findings in Sec. IV.

\section{Nondegenerate nonlinear dimer impurity}

We consider a system of one dimensional chain consisting of a nondegenerate
nonlinear dimer impurity of the kind $\chi |C|^{\sigma}$ (power law impurity)
where $\sigma$ is arbitrary. The dimeric impurity is placed at sites 0 and 1 
of the chain. The dimer is nondegenerate
in the sense that nonlinear strengths at sites 0 and 1 are different and
denoted by $\chi_0$ and $\chi_1$.
Therefore, the relevant Hamiltonian for this system is

\begin{equation}
H=\frac{1}{2} \sum_{m=-\infty}^{\infty} [C_m C_{m+1}^\star + h.c. ] +
\frac{V-1}{2} [C_0 C_1^\star + h.c. ] + \frac{\chi_0}{\sigma +2} 
|C_0|^{\sigma +2} +\frac{\chi_1}{\sigma +2} |C_1|^{\sigma+2}.
\end{equation}
Hopping matrix element between sites 0 and 1 is $V$ and others are taken to be
unity. The Hamiltonian with $V=1$ describes a perfect chain,
$V<1$ describes a system of perfect Cayley tree while $V>1$ describes
1- dimensional chain with a bond defect between sites 0 and 1. Furthermore, 
all the systems contain a nondegenerate nonlinear dimeric impurity
spanning the zeroth and first sites. In case of the Cayley tree $V$ takes
the form of $\frac{1}{\sqrt{K}}$, where $K$ stands for the connectivity of the
Cayley tree. The relevant transformation of the Cayley tree into an effective
1-dimensional system has been already presented [24].  
Since we are interested in the possible solutions for SL
states due to the presence of dimeric impurity we assume 
 
\begin{equation}
C_m=\phi_m e^{-iEt}
\end{equation}
The presence of a dimeric impurity in the system suggests us to consider the
 following form of $\phi_m$
\begin{eqnarray}
\phi_m=[Sgn(E)\eta]^{m-1} \phi_1    ~~~~~~   m \ge 1\nonumber\\
\phi_{-|m|}=[Sgn(E)\eta]^{|m|} \phi_0 ~~~~~~ m \le 0
\end{eqnarray}
We further note that the above form of $\phi_m$ 
can be derived from Green's function analysis [20].
Here $\eta \in [0,1]$ is given by $\eta = \frac{1}{2} [|E|-\sqrt{E^2-4}]$.
We further define $\beta = \phi_0/\phi_1 \in [-1, 1]$. 
Now normalization condition,$\sum_{m=-\infty}^{\infty}|C_m|^2=1$ together with
eqs.(3) and (4) give
\begin{equation}
|\phi_0|^2 = \frac{1-\eta^2}{1+\beta^2}
\end{equation}

Using eqs.(3), (4) and (5) in the Hamiltonian $H$, we obtain the
effective Hamiltonian of the reduced dynamical system given by
\begin{equation}
H_{eff}=Sgn(E) \eta + V \beta \frac{1-\eta^2}{1+\beta^2} + 
\frac{1}{\sigma +2} \left (\frac{1-\eta^2}{1+\beta^2}\right )^
{\frac{\sigma+2}{2}}
\left (\chi(1+|\beta|^{\sigma+2}) + \delta(1-|\beta|^{\sigma+2})\right )
\end{equation}
Here we have defined $\chi$ and $\delta$ as
$\chi=(\chi_0+\chi_1)/2$ and $\delta=(\chi_0-\chi_1)/2$. 
The Hamiltonian consists of two variables namely $\beta$ and $\eta$. 
SL states can be obtained from the fixed point solutions [21-23]
of the Hamiltonian, $H_{eff}$, of the reduced dynamical system. The fixed point
solutions satisfy equations $\partial{H_{eff}}/\partial{\beta}=0$ and
$\partial{H_{eff}}/\partial{\eta}=0$.

Equation $\partial{H_{eff}}/\partial{\beta}=0$ gives
\begin{equation}
[1-\eta]^{\frac{\sigma}{2}}=\frac{V(1-\beta^2)(1+\beta^2)^\frac{\sigma}{2}}
{\beta[(\chi+\delta)-|\beta|^{\sigma}(\chi-\delta)]}.
\end{equation}
Equation $\partial{H_{eff}}/\partial{\eta}=0$ gives
\begin{equation} 
\eta=\frac{Sgn(E) \beta [(\chi+\delta)-|\beta|^\sigma(\chi-\delta)]}
{V[(\chi+\delta)-|\beta|^{\sigma+2}(\chi-\delta)]}.
\end{equation}
For fixed values of $\chi$ and $\delta$ number of solutions satisfying 
eqs.(7) and (8) simultaneously will give the number of possible SL states.
This can be obtained for arbitrary $\sigma$ but we will consider 
$\sigma=0$ and $\sigma =2$. The reason behind taking $\sigma=0$
case is that the result is already known and hence the appropriateness of
the ansatz can be verified. $\sigma=2$ is considered 
because it is physically more relevant.

\noindent {\bf 1.  $\sigma=0$}

For $\sigma=0$, the system reduces to a perfectly linear 1-dimensional chain
with two static impurities $\chi_0$ and $\chi_1$ at sites 0 and 1 respectively.
In this case eq.(7) directly gives
\begin{equation}
\beta_{\pm}=-\frac{\delta}{V} \pm \sqrt{\frac{\delta^2}{V^2}+1}.
\end{equation}
In passing we note that , when $\delta=0$ i.e. the dimer impurity becomes
linear and degenerate, $\beta=\pm1$ are only permissible solutions. This
is consistent with our earlier work [20]. On the other hand $\delta\ne0$
necessitates $\beta\ne \pm1$ for permissible solutions.
Substituting eq.(9) in eq.(8) we obtain
\begin{equation}
\frac{1}{\chi}=\frac{\eta}{Sgn(E) \mp \eta\sqrt{\delta^2 + V^2}}.
\end{equation}
Eq.(10) along with signature of $E$ gives four
expressions for $1/\chi$. From each expression we will get a critical line
in the $(\chi, \delta )$ plane. The critical lines can be obtained by putting
$\eta =1$ in eq.(10). The equations describing the critical lines are
\begin{eqnarray}
\chi_c^{(1)}&=&1 + \sqrt{\delta^2 + V^2} \nonumber \\
\chi_c^{(2)}&=&-1 - \sqrt{\delta^2 + V^2} \nonumber \\
\chi_c^{(3)}&=&1 - \sqrt{\delta^2 + V^2} \nonumber \\
\chi_c^{(4)}&=&-1 + \sqrt{\delta^2 + V^2}. 
\end{eqnarray}
These critical lines are shown in fig.(1) for $V=1$, in fig.(2) for $V=0.5$
(which corresponds to a Cayley tree with $K$ = 4) and in fig.(3) for 
$V=\surd 2$. There are several regions in the $(\chi, \delta )$ plane 
bounded by the critical lines in all the figures. There are regions
containing no SL state, one SL state appearing above the host band, one
SL state appearing below the band, two SL states with one below and one above 
the band and two SL states appearing above as well as below the band.
Since our results for $V=1$ and $V=0.5$ agree with known results [25],
the validity of our starting ansatz is established.

\noindent {\bf 2.  $\sigma=2$}

We now investigate the case where nonlinearity power $\sigma=2$ .
Physically $\sigma=2$ can appear when the masses at the lattice points are 
treated as Einstein oscillators. From eq.(7) we again note that for 
$\delta=0$, $\beta=\pm1$ are permissible solutions. On the other hand, 
for $\delta\ne0$, $\beta=\pm1$ are not permissible solutions.
From eqs.(7) and (8) we obtain the relevant equation for SL states.
The equation is given as
\begin{equation}
\frac{1}{\chi}=\frac{\beta[(1+\alpha)-(1-\alpha)\beta^2]
\left[V^2 [(1+\alpha)-(1-\alpha)\beta^4]^2 -
 \beta^2[(1+\alpha)-(1-\alpha)\beta^2]^2 \right]}
{V^3 [1-\beta^4][(1+\alpha)-(1-\alpha)\beta^4]^2}
\end{equation}

where $\alpha=\delta/\chi$.
Using eq.(12) we have obtained the full phase diagram of SL states in 
$(\chi,\delta)$ plane for $V=1$ and this is shown in fig.(4). There are 
various regions in the phase diagram numbered according to number of 
possible SL states in the region. Along the $\delta = 0$ line, we note 
that for $\chi\ge 0$, there are three critical values of 
$\chi$, namely, 1, 8/3 and 8. These critical values are in agreement with 
the values obtained in our earlier work [23]. In this case, maximum 
number of SL states appearing is five. With the use of same eq.(12), we
have also obtained the phase diagram for SL states in $(\chi,\delta)$ plane
for a Cayley tree with connectivity $K=4$ $(V=0.5)$. This is shown in fig.(5).
Maximum number of possible SL states in this case is found to be seven.
The critical values of $\chi$ along $\delta=0$ line are again in agreement 
with our earlier result [24].
Typical phase diagram of SL states in the $(\chi, \delta)$ plane with 
$V=2$ is also shown in fig.(6). The phase diagram contains one to seven 
SL states. Furthermore, if we compare the results for $\delta = 0$ with 
$\delta \ne 0$, we note that maximum number of possible SL states increases
as nondegeneracy is introduced.

In passing it is worth to mention that there are $N/2$ stable SL states
in a region if the region in the phase plane contain $N$ number of SL states 
and $N$ is even. On the other hand $(N+1)/2$ states are stable if $N$ is odd.
The stability of a SL state is connected with the variation of the energy of 
the state as a function of $\chi$. If the energy of the state increases with 
the increase of $\chi$, the state is stable otherwise unstable. A simpler 
analysis of stability of a state and its connection with its energy will be 
presented in the next section.

\section{Fully non-linear chain with alternative nonlinear strengths}

Here we consider a fully non-linear 1-dimensional system where the alternative
sites are of different strengths. The Hamiltonian of the system is given by
\begin{equation}
H=\frac{1}{\sigma+2} \sum_{n=-\infty}^{\infty} \chi_n |C_n|^{\sigma+2}
+\frac{1}{2} \sum_{n=-\infty}^{\infty} [C_n^\star C_{n+1} + h.c.]
\end{equation}
where ,
\begin{eqnarray}
\chi_{2n} &=& \chi_1,~~~-\infty \le n \le \infty \nonumber\\
\chi_{2n+1} &=& \chi_2,~~~-\infty \le n \le\infty .
\end{eqnarray} 
Hopping matrix elements between neighboring sites are taken to be unity.
To obtain SL states we consider $C_n = \phi_n e ^{-iEt}$  and
$\phi_n=\phi_0 \eta ^{|n|}$ . The expression for $\eta$ is obtained by
taking $|n| \rightarrow \infty$ and is given by $\eta = \frac{1}{2} [|E|-
\sqrt{E^2-4}]$ [21]. Here we have
considered monomeric ansatz because the system is symmetric about the 0-th
site. After going through the same procedure as earlier and using the
normalization condition , $\phi_0 ^2 = (1-\eta^2) /2$ ,
we obtain the reduced Hamiltonian of the dynamical system given by
\begin{equation}
H_{eff}=\frac{1}{\chi(\sigma+2)} \left[\frac{1-\eta^2}{1+\eta^2}\right]^
\frac{\sigma+2}{2} \left [\chi \frac {1+\eta^{\sigma+2}}{1-\eta^{\sigma+2}} +
\delta \frac {1-\eta^{\sigma+2}}{1+\eta^{\sigma+2}} \right]
+ \frac{2\eta}{1+\eta^2}
\end{equation}
where $\chi$ and $\delta$ are defined as $\chi=(\chi_1 + \chi_2)/2$ and
$\delta=(\chi_1 - \chi_2)/2$.
Setting $\partial{H_{eff}}/\partial{\eta}=0$ we get 
\begin{equation}
\frac{1}{\chi}=\frac{(\eta^\sigma-1)(\eta^{\sigma+4}+1)}
{\left[\frac{\eta^2-1}{\eta} \left(\frac{1+\eta^2}{1-\eta^2}\right)^
\frac{\sigma}{2} -\delta\frac{(\eta^\sigma+1)(\eta^{\sigma+4}-1)}{(1+
\eta^{\sigma+2})^2}\right] (1-\eta^{\sigma+2})^2}.
\end{equation}
    
Eq.(16) can be used to analyze number of possible states for different values 
of $\chi , \delta {\rm~and~} \sigma$. For $\sigma=0$ and $\delta=0$ from 
eq.(16) we
see that to get a SL state $\chi$ needs to be infinite. It, therefore, means 
that no SL states can be obtained. This is true for $\sigma=0$ and $\delta=0$
because the system reduces to a perfect linear system. This is also true for
$\sigma=0$ and $\delta\ne0$.  Analysis of SL states can be 
done for arbitrary $\sigma$ but we consider the case of $\sigma=2$ for
the reason mentioned earlier. For $\sigma=2$, eq.(16) reduces to
\begin{equation}
\frac{1}{\chi}=\frac{\eta(1+\eta^6)(1+\eta^4)^2}{(1+\eta^2)^3
[(1-\eta^2)(1+\eta^4)^2 - \delta \eta (1-\eta^2)(1-\eta^6)]}  
\end{equation}
Eq.(17) directly tells that for $\delta=0$ there will be always one SL state
and this is consistent with our earlier result [24]. Using eq.(17) 
the phase diagram of SL states in $(\chi,\delta)$  plane is obtained and 
shown in fig.(7).  As usual numbers in different regions indicate number of 
SL states in those regions. Here also maximum number of SL states increases 
compared to a perfect nonlinear chain. Furthermore the phase 
diagram in this case is quite rich (see fig.(7)).

The stability of the SL states in this system can be understood from a simpler 
graphical analysis. For this purpose we look at the Hamiltonian given in
eq.(15). The Hamiltonian is function of one dynamical variable, namely, $\eta$
where $\chi$ and $\delta$ are parameters. For fixed values of $\chi$ and 
$\delta$, $\eta$ and $\partial{H_eff}/\partial{\eta}=f(\eta)$ can be treated 
as generalized coordinate and generalized momentum respectively for the reduced dynamical system. 
Solutions of the equation $f(\eta)=0$ will give the fixed points for fixed 
value of $\chi$ and $\delta$ and number of fixed points with $\eta \in [0,1]$ 
will be the number of SL states. We now plot $f(\eta)$ as a function of
$\eta$ for $\chi=0.35$ and $\delta=2.5$ as shown in fig.(8). If $f(\eta)
> 0$, then the flow of the dynamical variable will be along the positive 
direction and on the other hand if $f(\eta) < 0$ the flow will be along the 
negative direction. These flows are indicated by arrows in the figure and the 
fixed points are denoted by A, B and C respectively. In the neighborhood of 
A the flow is always towards A and same for C. Therefore, 
A and C are stable fixed points. On the other hand in the neighborhood 
of B the flow is away from B. Hence, this is an unstable 
fixed point. Same thing happens for all $(\chi,\delta)$ in the three state 
region in fig.(7). We therefore notice that among three SL states (fixed 
points) two are stable and the other one is unstable. To get a connection of 
the stability of the states with the energy and $\chi$, we plot the energy of 
the states as a function of $\chi$ in the neighborhood of $\chi=0.35$ for 
fixed value of $\delta =2.5$ in fig.(9). Energies of the SL states arising 
from the fixed points A, B and C are denoted by $A_1$, $B_1$ and $C_1$
respectively in fig.(9). The figure clearly shows that in the neighborhood
of $\chi=0.35$, the energy of two stable SL states increases with $\chi$
and that of the unstable state decreases. We, therefore, conclude that the 
SL state is stable if the energy of the state increases with $\chi$ otherwise 
unstable.

\section{summary}

We have found the possible number of SL states due to the presence of 
a quadratic nondegenerate nonlinear dimeric impurity in one dimensional
systems. Phase diagram of SL states in $(\chi,\delta)$ plane is presented
for all the systems with a nondegenerate dimeric impurity. Maximum number of SL states is 
found to increase due to the introduction of nondegeneracy. The dimeric ansatz
has been used in this case and this reproduces known results for special cases.
A full phase diagram of SL states for a system of one dimensional chain
with alternative sites having different nonlinear strength is presented.
A monomeric ansatz has been introduced in this case
and found to be consistent for special cases. Here also maximum number of SL
states increases compared to that in a perfect nonlinear system.
 A stability analysis for
the SL states in the fully nonlinear system and a connection of the
stability with the variation of the energy of the state as a function 
of the nonlinear parameter are presented.

\begin{figure}
\caption{The phase diagram of SL states in $(\chi,\delta)$ plane
for an one dimensional system with a nondegenerate linear dimeric
impurity in the middle of the system. The $(\chi ,\delta)$ plane is
divided into many regions by the critical lines. The region indicated
by 2LA implied the possibility of two SL states in that region and both
of them lie above the band of the linear host. Similarly 1LA, 1LB and
2LB implies one SL state above the band, one SL state below the band
and two SL states below the band respectively.}
\end{figure}
\begin{figure}
\caption{The phase diagram of SL states in $(\chi,\delta)$ plane for a 
Cayley tree with a nondegenerate linear dimeric impurity embedded in 
the middle of the system. The connectivity ($K$) of the Cayley tree is 4. 
The region indicated by OL implies that there is no possible SL state 
in that region. Others are same as in fig. (1).}
\end{figure}
\begin{figure}
\caption{The phase diagram of SL states in $(\chi,\delta)$ plane for
one dimensional system with a bond defect $(V=\surd 2)$ between sites 0 and 
1 as well
as a nondegenerate linear dimeric impurity occupying sites 0 and 1. 
Other things are same as fig. (1) and fig. (2).}
\end{figure}
\begin{figure}
\caption{The phase diagram of SL states in $(\chi,\delta)$ plane
for an one dimensional system with a nondegenerate quadratic nonlinear 
dimeric impurity in the middle of the system. The $(\chi,\delta)$ plane is
divided into many regions by the critical lines. The number in a region
indicate the number of possible SL states in the region.}
\end{figure}
\begin{figure}
\caption{The phase diagram of SL states in $(\chi,\delta)$ plane for a
Cayley tree with a nondegenerate quadratic nonlinear dimeric impurity 
embedded in the middle of the system. Here $K$ = 4.
The number in a region indicates the number of possible SL states 
in the region. There are two small regions near the origin containing two 
SL states.}
\end{figure}
\begin{figure}
\caption{The phase diagram of SL states in $(\chi,\delta)$ plane for
one dimensional system with a bond defect between sites 0 and 1 as well
as a nondegenerate quadratic nonlinear dimeric impurity occupying sites 
0 and 1. The bond defect, $V$ is taken to be 2. The number in
a region indicates the number of possible SL states in the region.}
\end{figure}
\begin{figure}
\caption{Phase diagram of SL states in $(\chi,\delta)$ plane for a fully 
nonlinear one dimensional system where alternative sites are of different
nonlinear strengths. The number in a region indicates the number of possible 
SL states in the region.}
\end{figure}
\begin{figure}
\caption{$f(\eta)$ is plotted as a function of $\eta$. Points indicated by A, 
B and C are fixed points. Here $\chi$ and $\delta$ are taken to be 0.35 and
2.5 respectively.}
\end{figure}
\begin{figure}
\caption{Energy of SL states are plotted as a function of $\chi$. Here 
$\delta$ = 2.5. $A_1$, $B_1$ and $C_1$ are the points corresponding to
the points A, B and C ( in fig. (8) ) respectively. The dotted line is
$\chi=0.35$. }
\end{figure}
\end{document}